\documentclass[conference]{IEEEtran}
\pdfoutput=1
\IEEEoverridecommandlockouts

\usepackage{url}
\usepackage{graphicx}
\usepackage{fixme}
\usepackage{url}

\begin{document}
%
\title{Dependability in a Multi-tenant Multi-framework Deep Learning as-a-Service Platform\thanks{Authors' names listed in alphabetical order. The authors would like to thank Khoa Hyunh of IBM for his help evaluating DLaaS performance overhead.}}


\author{\IEEEauthorblockN{Scott Boag, Parijat Dube, Kaoutar El Maghraoui, Benjamin Herta, Waldemar Hummer, \\ K. R. Jayaram, Rania Khalaf, Vinod Muthusamy, Michael Kalantar, Archit Verma}
\IEEEauthorblockA{IBM Research AI\\
\emph{\{scott\_boag,pdube,kelmaghr,bherta,whummer,jayaramkr,rkhalaf,vmuthus,mkalantar,archit.verma\}@us.ibm.com}}}

\maketitle


\begin{abstract}
Deep learning (DL), a form of machine learning, is becoming increasingly popular in several application domains. As a result, cloud-based Deep Learning as a Service (DLaaS) platforms have become an essential infrastructure in many organizations. These systems  accept, schedule, manage and execute DL training jobs at scale.  

This paper explores \emph{dependability} in the context of a DLaaS platform used in IBM. We begin by explaining how DL training workloads are different, and what features ensure dependability in this context. We then describe the architecture, design and implementation of a cloud-based orchestration system for DL training. We show how this system has been architected with dependability in mind while also being horizontally scalable, elastic, flexible and efficient. We also present an initial empirical evaluation of the overheads introduced by our platform, and discuss tradeoffs between efficiency and dependability.
\end{abstract}


\section{Motivation and Introduction}~\label{sec:introduction}

The increasing popularity of deep learning can be attributed
to the following four factors: 
(1) Artificial neural networks can often learn features in an unsupervised manner, taking
feature engineering out of the picture, (2) Recent improvements in GPU technologies 
have made large scale matrix computations
typical of deep-learning algorithms effective, (3) Advances in interconnection 
technologies and data center networking technologies like NVLink, Infiniband and 100G
Ethernet have enabled \emph{distributed} DL training algorithms to effectively 
synchronize by transferring large amounts of training data and models, and (4)
widely available \emph{open-source} deep learning frameworks like Tensorflow, PyTorch, 
Caffe, Torch, Theano, Horovod and MXNet have reduced the effort required to design, train, and use deep learning models.

While advances in hardware have enabled DL to scale, said hardware remains expensive 
and should be effectively utilized to obtain good returns on investment. 
A cloud-based distributed deep learning \emph{platform} helps organizations (like ours)
utilize expensive hardware effectively, and enables developers, applications and
customers to share deep learning infrastructure. IBM Deep Learning as a Service (DLaaS) is
a distributed cloud-based software platform that handles the scheduling, 
orchestration, elasticity and resilience of deep learning jobs, and is agnostic 
to the internals of the deep learning job. DLaaS aims to reduce the barrier 
to entry \emph{even further} by enabling developers to focus on training neural nets 
and choosing hyper-parameters rather than focusing on installation, configuration and 
fault tolerance. 

DLaaS has four main goals -- (1) \emph{Flexibility} to support different deep-learning
frameworks. (2) \emph{Scalability}, i.e., horizontal scalability or the ability to 
manage increasing numbers of deep learning jobs 
by increasing the hardware resources available to the platform, (3) \emph{Dependability}, 
meaning that the platform should be highly available, reliable and handle faults
in a robust manner, secure and maintainable, and (4) \emph{Efficiency}, meaning that the 
overheads introduced by the platform to achieve aforementioned goals (especially flexibility 
and dependability) and the response time of the platform to external requests must be minimal. The goal of this paper is to examine \emph{dependability} in the context of IBM DLaaS.

\section{Dependability Challenges in Deep Learning}\label{sec:challenges}

DL training jobs have unique characteristics, which introduce 
dependability challenges in DLaaS:

\begin{itemize}
    \item DL training jobs are typically run for a few days (1-7) continuously, for hundreds of thousands of iterations over a large data set. So the consequences of failure can be large 
    (potential loss of several days of work).
    
    \item DL jobs are GPU-heavy, and are engineered to exploit the massive SIMD parallelization in GPUs and maximize GPU utilization. This increases heat generated by GPU servers in the datacenter, and server machine failures (typically reboots, power downs, etc.) are not uncommon.
    
    \item DL jobs impose a heavier load on datacenter networks. DL algorithms make several passes over the data set, which can be tens (or sometimes hundreds) of TB. At these sizes, data cannot be 
    stored locally and typically has to be streamed over the network (either from a cloud-based 
    store or NFS) for \emph{each} pass. 
    
\end{itemize}

In addition, operating a flexible multi-framework, multi-tenant deep-learning as a service 
platform supporting single node and distributed DL jobs 
requires the following dependability guarantees:

\begin{itemize}
    \item Deploying a DL job is seldom instantaneous; it is a multi-step process, involving placement 
    on an appropriate cluster node with available GPUs, setting up network (MPI) interconnections,
    provisioning shared volumes and credentials to access data, etc. Users require that provisioning of
    DL jobs be atomic, either the whole job is provisioned with the requisite resources or none.

    \item Given that DL jobs are long running, users expect periodic and accurate status updates 
    (e.g., whether the job is DEPLOYING, PROCESSING). These status updates should be dependable because
    users use associated timestamps for job profiling and debugging.
    
    \item Reliable streaming of logs from the job, irrespective of the stage it is in, even if
    it crashes/fails. This is key for users to debug their jobs.
    
    \item DL frameworks are so flexible that e.g., a Tensorflow job can be arbitrary customer 
    code. Hence, for multi-tenancy, DL jobs must be isolated from DLaaS system processes, and from
    each other.
    
    \item Support for both user-directed and configurable automatic periodic checkpointing,
    given the longevity of DL jobs.
    
    \item Resilience to node and job crashes. Both failure detection and 
    recovery are important because users expect to be notified
    when DL jobs are restarted, because ``training progress graphs'' differ (slightly) between
    a job that never experienced a failure and a job that did.
    
\end{itemize}

\section{Dependability in IBM DLAAS}\label{sec:dlaas}

DLaaS is a cloud-native application architected as a set of loosely-coupled 
microservices communicating with each other using GRPC. Logically, DLaaS has three
layers (1) DLaaS Core-Services Layer, consisting of two components/microservices -- 
API and Lifecycle Manager (LCM) (2) DLaaS Platform Layer
which consists of the infrastructure that the core-services rely on -- Docker, Kubernetes~\cite{kubernetes}, 
ETCD~\cite{etcd}, and MongoDB~\cite{mongodb}, and (3) DLaaS Helpers -- components which are part of the DL job during execution.
Helpers perform failure detection, status recording and updates, log collection, data/results
transfer, and metrics collection.

\begin{figure}
    \centering
    \includegraphics[width=\columnwidth]{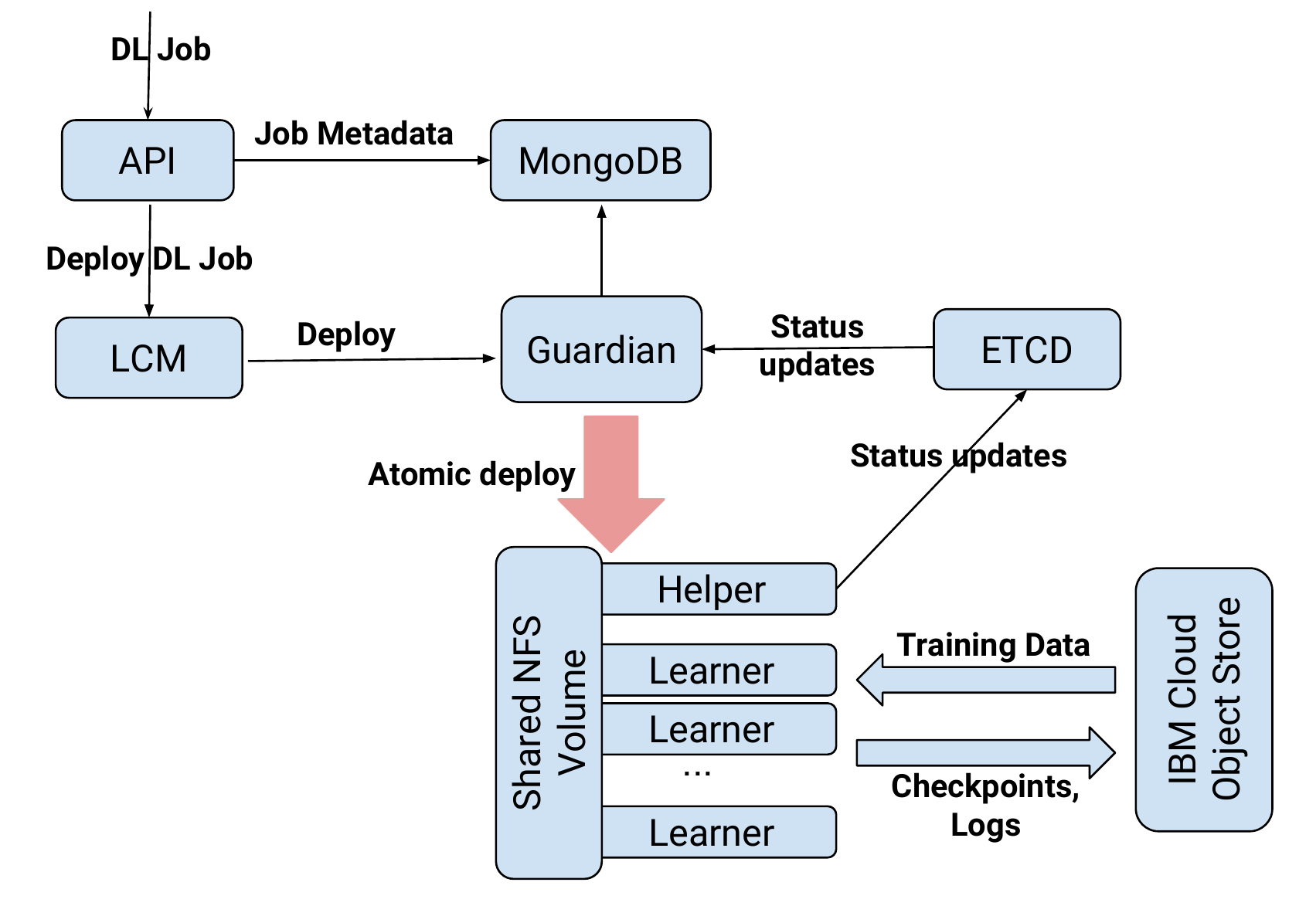}
    \caption{DLaaS Core Services and Training Jobs}
    \label{fig:dlaasjobflow}
\end{figure}

\paragraph{A DL Training Job} DLaaS supports several popular DL frameworks like Caffe, Torch, 
Horovod, etc. DLaaS maintains Docker images corresponding to each of these frameworks. In its simplest
form, a DL training job consists of a single learning process (``learner'') in a Docker container 
using a GPU, i.e., a framework docker image instantiated with user code. Typically, DL jobs
use several GPUs and/or consist of several learners synchronizing over MPI or using a 
centralized parameter server. Users submit training jobs and manage them using 
the DLaaS API (both GRPC and REST are supported). Job parameters, including the source of training data,
credentials to access training data, framework, number of learners, 
location where results and logs should be stored, learning rate,
etc., are specified using a manifest file. 

\paragraph{Cluster Management} DLaaS employs Kubernetes (K8S)~\cite{kubernetes} for container orchestration and 
cluster management. A K8S \emph{pod}
is a group of one or more containers (such as Docker containers), with shared storage/network, 
and a specification for how to run the containers. A pod's contents are always 
co-located and co-scheduled, and run in a shared context. 
All containerized DLaaS core services are executed as 
K8S \emph{deployments}, exposed through the K8S \emph{service} abstraction.
DLaaS core services use K8S to deploy containerized DL jobs using appropriate
K8S abstractions (Jobs  and Stateful Sets).

\paragraph{DLaaS Core Services} The DLaaS API microservice handles all the incoming 
API requests including load balancing, metering, and access management. It exposes both a
RESTful API as well as a GRPC API endpoint. The API service instances are dynamically registered
into a K8S service registry that provides load balancing and fail-over support 
for incoming API requests. For the lifetime of a DL job, all its metadata, including its 
job parameters, are stored in MongoDB~\cite{mongodb}. When a job deployment request arrives, the API layer
stores all the metadata in MongoDB before acknowledging the request. 
This ensures that submitted jobs are never lost.
The API layer then submits the job to the DLaaS Lifecycle Manager (LCM) microservice. 
As its name suggests, the LCM is responsible for the job from submission to 
completion/failure, i.e., the deployment, monitoring, garbage collection, and user-initiated
termination of the job.

\paragraph{Atomic Job Deployment} The LCM uses a Kubernetes (K8S) abstraction (unfortunately also
called a ``Job'') for atomic deployment of DL jobs. K8S Jobs are essentially tasks (i.e. Docker containerized
code) that K8S guarantees to reliably run to completion \emph{exactly once}. If a K8S Job crashes for
any reason (like a OS, Docker, K8S or machine failure), K8S will automatically restart it and 
execute it again. To deploy a DL job, the LCM simply instantiates a component called the 
\emph{Guardian} with all the metadata of the DL job. The Guardian is a DLaaS component created
on the fly as a K8S Job for every DL job. Creation of the Guardian is a very quick (less than 3s
in our experiments) single step process. The Guardian then executes the multi-step process
of actually deploying the DL job by further interacting with K8S. This involves instantiating
Docker containers (corresponding the DL framework used, like Caffe. Torch, etc.)
with training parameters and user code,
setting up shared NFS volumes to monitor training progress, K8S policies to restrict network
access from the learner in a multi-tenant environment, etc. If the Guardian crashes in the middle
of a job deployment, K8S is guaranteed to restart it. The restarted Guardian will \emph{roll back}
the previous partially deployed DL job and starts a fresh deployment process. In the presence of
persistent failures, this process will be repeated for a (configurable) number of times
before the Guardian gives up and marks the DL job in MongoDB as FAILED. Once a DL job is successfully deployed, the Guardian is then responsible for
monitoring its progress.

\paragraph{Detecting Failure/Completion of Learner Processes} The Guardian uses the K8S abstraction called
\emph{Stateful Set} to deploy a DL Job. This enables DLaaS to create replicated 
learners (for distributed training) and is well suited for DL frameworks like Horovod and
distributed Tensorflow. For each DL job, the Guardian also creates a separate \emph{helper K8S pod} using the K8S
\emph{Deployment} abstraction, which contains a number of ``helper'' containers -- load-data,
log collector, store-results, and controller. The helper pod remains isolated from the learner pods, 
but both share a common NFS filesystem, mounted by the Guardian using a K8S persistent volume claim.
The shared NFS volume enables the controller container running separately in the helper pod 
to monitor the execution and exit
status of the learner processes and detect both learner process completion and 
failures by reading their output (e.g.,exit status redirected to a file). 

\paragraph{Reliable Status Updates} In addition to detecting completion and failure, the controller 
 can read the status/output of the load-data and store-results containers
because all the helper and learner containers share a common file system. To reduce coupling between DLaaS components
and ensure reliable 
status updates, we employ the ETCD key-value store~\cite{etcd} to co-ordinate between the controller and LCM/Guardian.
ETCD itself is replicated (3-way), and uses the Raft consensus protocol to ensure consistency.
The controller records the current status of each learner in ETCD, where it is read by the Guardian.
The Guardian aggregates the statuses of each learner to record the overall status of the job in
MongoDB, from where the user can read it through a REST/GRPC API call to DLaaS. Using ETCD makes 
status updates resilient to crashes of both the controller/helper pod and crashes of the Guardian. Using NFS makes status updates resilient to controller crashes; K8S will restart the
controller which can read current status and previous statuses from NFS. 

\paragraph{Checkpointing} Given the long running nature of DL training jobs, checkpointing is 
vital. DLaaS enables users to configure checkpointing intervals; and checkpoints are stored in
a cloud-hosted object store. The checkpointing interval depends on the tolerance level of the user
to failures, i.e., how many hours of work the user is willing to lose in the event of a failure.
Typically, users execute training jobs on a local laptop/server
on a small subset of the input to profile the training job and identify good checkpointing intervals;
and specify these intervals as parameters while submitting the DL job.

\paragraph{Node/Container Crashes} Orderly learner failures, i.e., by writing 
an appropriate exit code to NFS, can be detected by the controller. However, DL
job \emph{crashes} due to node/container crashes are handled by K8S. Crashed learners will be restarted automatically by K8S, because learners are deployed as stateful sets. A recovered learner
can continue training either (1) from the latest checkpoint (2) in the case of distributed DL jobs,
by rejoining other learners and getting the latest neural net parameters from a parameter server (if the
DL framework supports this). The amount of work lost due to a crash is determined by the checkpointing interval.

\section{Evaluation}~\label{sec:evaluation}

In this section, we demonstrate empirically that the dependability features of 
DLaaS and execution in a containerized environment have minimal impact on performance
of DL training jobs. We illustrate this by using several DL benchmarks~\cite{dlbenchmarks}
(VGG-16~\cite{vgg16}, Resnet-50~\cite{resnet50} and InceptionV3~\cite{inceptionv3}),
two different PCIe-based GPU types (K80~\cite{k80} and P100~\cite{p100}), and 
two different DL frameworks (Caffe v1.0~\cite{caffe} and TensorFlow v1.5~\cite{tensorflow}).

\begin{figure}[htb]
    \begin{tabular}{c|c|c|c}
    \hline
    Benchmark   & Framework  &  \# PCIe  & Difference in  \\
                &            &     K80 GPUs  & Performance \\
    \hline
    VGG-16  & Caffe & 1 & 3.29\% \\
    \hline
    VGG-16 & Caffe & 2 &  0.34\% \\
    \hline
    VGG-16 & Caffe & 3 &  5.88\% \\
    \hline
    VGG-16 & Caffe & 4 &  5.2\% \\
    \hline
    InceptionV3 & TensorFlow & 1 & 0.32\% \\
    \hline
        InceptionV3 & TensorFlow & 2 & 4.86\% \\
    \hline
        InceptionV3 & TensorFlow & 3 & 5.15\% \\
    \hline
        InceptionV3 & TensorFlow & 4 & 1.54\% \\
    \hline
    \end{tabular}
    \caption{Performance overhead of DLaaS vs. IBM Cloud Bare Metal Servers on popular Image Processing Benchmarks. Performance is quantified as images processed/sec for training. Caffe v1.0 and Tensorflow v1.5 were used.}\label{fig:overhead-sl}
\end{figure}

For our first set of measurements, we compare DLaaS deployed on IBM Cloud
with directly executing the benchmarks (non containerized) on bare metal machines manually
on IBM Cloud datacenters. 1GbE interconnect was used in both cases. Training data was
stored in IBM Cloud Object Store. Results are illustrated in Figure~\ref{fig:overhead-sl}.
From Figure~\ref{fig:overhead-sl}, we observe that performance overhead induced by DLaaS is 
minimal (when compared to dependability and ease of use)

\begin{figure}
    \centering
    \begin{tabular}{c|c|c|c|c}
    \hline
    Benchmark   & Framework  &  \# PCIe  & GPU & Difference in  \\
                &            &    GPUs  & Type & Performance \\
    \hline
    Inceptionv3 & TensorFlow & 1 & P100 & 3.30\% \\
    \hline
    Resnet-50 & TensorFlow & 1 & P100 & 7.07\%  \\
        \hline
    VGG-16 & TensorFlow & 1 & P100 & 7.84\% \\
    \hline
    InceptionV3 & Tensorflow & 2 & P100 & 10.06\% \\
    \hline
    Resnet-50 & Tensorflow & 2 & P100 & 10.53\% \\
        \hline
    VGG-16 & Tensorflow & 2 & P100 & 13.69\% \\
    \hline
    \end{tabular}
    \caption{Performance overhead of DLaaS vs. NVidia DGX-1 bare metal server on TensorFlow HPM benchmarks~\cite{tensorflowcnn}. Performance is quantified as images processed/sec for training. }
    \label{fig:overhead-dgx}
\end{figure}

For the second set of measurements, we compare DLaaS to NVidia's specialized 
hardware -- DGX-1~\cite{dgx}, which incurs $\approx$2-3$\times$ in additional 
costs compared to off-the-shelf hardware (such as IBM Cloud~\cite{ibmcloudgpu}).
DGX-1 has advanced hardware (NVLink and High Bandwidth Memory), and is expected
to have higher performance than DLaaS. However, we observe from Figure~\ref{fig:overhead-dgx},
that degradation in performance, though nontrivial is only modest (up to $\approx$ 15\%).

\begin{figure}
    \centering
    \begin{tabular}{c|c}
    Component & Time to recover \\
              & from crash failure \\
              
    \hline
    API  & 3-5s \\
    LCM  & 4-6s \\
    Guardian & 1-2s \\
    Helper & 3-4s \\
    Learner & 10-20s \\
    \end{tabular}
    \caption{Time taken to recover from crash failures, by component.}
    \label{fig:recoverytime}
\end{figure}

Finally, our cloud-native design and implementation has ensured that 
DLaaS remains loosely coupled and each component can fail independently of the other. 
Within a DL training job, a learner can crash and be restarted by K8S
independently of the helper. Guardians can crash/restart independently of the LCM and API, and
so on. Time taken for each component to restart is minimal and illustrated in
Figuire~\ref{fig:recoverytime}. These times were calculated by manually crashing 
various components (using the kubectl tool of K8S~\cite{k8s}) and measuring 
time taken for the component to restart. Learners take longest to restart because binding to cloud object store
and persistent NFS volumes takes longer, and Caffe/Tensorflow pods take
longer to restart when compared to DLaaS microservice pods (written in GoLang).

\section{Related Work}
\label{sec:related}

Efforts to develop machine learning (ML) systems have appeared in industry and academia. Representatives of the former
include IBM Watson Machine Learning~\cite{ibm-wml}, Amazon SageMaker~\cite{amazon-sagemaker}, Google Cloud Machine
Learning~\cite{google-cloud}, and Microsoft Azure~\cite{microsoft-azure}. These offerings differ in their capabilities,
but none address the complete AI lifecycle and dependability issues of the underlying platform.
Li et al.~\cite{li:17} discuss challenges associated with building a scalable ML service, including feature computation
over global data. Their focus is mainly on real-time serving of large number of~models,
without considering the integration lifecycle. ModelHub~\cite{miao:17} and
ModelDB~\cite{vartak:16} are lifecycle management systems for ML models supporting efficient storing,
querying, and sharing of artifacts. These systems are focused on the model lifecycle, and do not consider the
co-evolution of the applications or platform optimizations.
While there has been a lot of focus on securing multi-tenant services~\cite{cloudsecurity} , there has been little
attention paid to DL workloads in such an environment.

\section{Conclusions}

This paper provides a \emph{brief} overview of how dependability is
addressed in the context of IBM DLaaS -- a publicly available, multi-tenant, multi-framework,
deep-learning as a service platform. It provides an initial evaluation of the 
efficacy of our approach, by measuring performance overhead. We have open-sourced
major portions of this platform at \cite{ffdl}, and hope it can be a foundation for
further research in this area.

\bibliographystyle{plain}
\bibliography{dsn2018}

\end{document}